\documentclass{article}

\usepackage[a4paper, width=170mm, top=25mm, bottom=25mm]{geometry}
\usepackage[]{graphicx}
\usepackage{color, colortbl}
\usepackage{amsmath,amsfonts,amsthm,bm}
\usepackage{xcolor}
\usepackage{float}
\usepackage{multicol} 
\usepackage{multirow} 
\usepackage{mathtools}
\usepackage{caption}
\usepackage{subcaption}
\usepackage{tabularx}
\usepackage{array,multirow} 
\usepackage{booktabs}
\usepackage{url}
\usepackage[colorinlistoftodos]{todonotes}
\usepackage{rotating}
\usepackage{hyperref}

\usepackage[
  sorting=none
]{biblatex}
\newcommand{\ms}{/\allowbreak}

\begin{document}


\title{Engineering T7 RNA Polymerase for High-Purity \\\textit{In Vitro} Transcription\\}
\author{Pauline Hermans$^{1,2,3}$,  Youlia Serikova$^{3}$, José Castillo$^{3}$, Marianne Rooman$^{1,2,\dagger}$,\\\, Fabrizio Pucci$^{1,2,\dagger}$ \\
 $^1$Computational Biology and Bioinformatics, Université Libre de Bruxelles, 1050 Brussels, Belgium\\
$^2$ Interuniversity Institute of Bioinformatics in Brussels, 1050 Brussels, Belgium\\
$^3$ Phoenix Biosciences S.A. d/b/a Quantoom Biosciences, Nivelles, Belgium
}
\date{\today}

\maketitle


\section*{Abstract}
\noindent

\textit{In vitro} transcription using bacteriophage T7 RNA polymerase (T7~RNAP) is the gold-standard platform for RNA production in both research and therapeutic applications. Despite its high processivity and promoter specificity, T7~RNAP generates multiple RNA by-products, including double-stranded RNA, 3'-extended transcripts, abortive RNAs, and prematurely terminated products. These impurities reduce RNA yield, complicate downstream purification, and raise safety concerns for RNA-based therapeutics by activating adverse innate immune pathways.
Although reaction optimization and downstream purification strategies can mitigate these issues, they typically involve trade-offs between RNA purity and yield. Enzyme engineering has therefore emerged as a powerful upstream strategy to suppress by-product formation at its molecular origin. 
Here, we synthesize current knowledge on the structural and mechanistic basis of T7~RNAP by-product formation and systematically review engineering strategies to improve RNA purity. T7~RNAP variants are classified according  to their underlying mechanisms of action, including enhanced thermostability, reduced non-specific template binding, smoother initiation-to-elongation transition, reduced premature termination, and template-biased polymerase designs.
This analysis identifies general principles governing the trade-off between specificity and processivity  and highlights synergistic combinations of mutations that improve RNA purity without compromising transcriptional efficiency. We conclude by discussing the remaining challenges for engineering T7~RNAP to meet the stringent purity requirements of next-generation RNA therapeutics.


\section{Introduction}

\textit{In vitro} transcription (IVT) using bacteriophage T7 RNA polymerase (T7~RNAP) has become a fundamental technology in molecular biology and biotechnology, enabling the synthesis of RNA molecules for a broad range of therapeutic applications, spanning immunogenic approaches that stimulate the immune system, such as prophylactic and cancer vaccines, as well as non-immunogenic approaches, including targeted genome editing and protein replacement therapies \cite{shi2024}.

T7~RNAP is a single-subunit enzyme encoded by gene~1 of bacteriophage~T7 and is therefore structurally simpler than the multi-subunit RNA polymerases found in bacteria and eukaryotes \cite{sousa2003, tunitskaya2002}. Its high processivity and strong promoter specificity enable the efficient production of RNA transcripts extending to several thousand nucleotides in length, establishing T7~RNAP as the gold-standard enzyme for IVT. However, T7~RNAP IVT frequently generates unwanted by-products, including double-stranded RNA (dsRNA) and truncated transcripts. These arise from diverse mechanisms such as non-template nucleotide addition, premature termination, transcriptional pausing, and RNA-dependent RNA polymerase (RdRP) activity \cite{jo2025, wu2020, he2024, nowak2026, martin1988}.

These impurities not only reduce RNA yield and necessitate costly downstream purification but also pose significant safety concerns in therapeutic applications. In particular, dsRNA and aberrant RNA species can activate innate immune sensors such as Toll-like receptors and protein kinase R, which trigger downstream immune signaling pathways and induce inflammatory responses. These responses can reduce RNA stability, impair translation efficiency, and ultimately diminish therapeutic efficacy \cite{jo2025, chen2022, roers2016}. For non-immunogenic RNA therapeutics, these effects are especially problematic, as achieving therapeutic benefit often requires protein expression levels up to 1000-fold higher than those needed for vaccines, potentially leading to exacerbated innate immune activation \cite{yu2024}.

Although substantial progress has been made in downstream purification technologies, upstream solutions that suppress impurity formation at its source offer a more scalable and economical strategy to suppress by-product formation \cite{miller2024}.
The structural simplicity of T7~RNAP, which comprises three distinct domains responsible for promoter recognition, nucleotide binding, and catalysis, makes it an attractive target for rational enzyme engineering aimed at improving transcriptional fidelity \cite{tunitskaya2002}. 

The literature on engineered T7~RNAP variants remains fragmented. Reported strategies, ranging from single-point mutations to multi-site variants and fusion constructs, are often evaluated under heterogeneous experimental conditions and without direct comparison of their mechanistic rationales or outcomes \cite{nowak2026}. This heterogeneity hinders the establishment of general design principles for robustly improving RNA purity across diverse IVT contexts.

To address this shortcoming, this paper reviews and integrates current knowledge on T7~RNAP structure, transcriptional mechanisms, and by-product formation pathways, with a particular focus on enzyme engineering strategies aimed at reducing dsRNA and size-variant RNA formation during IVT. By integrating structural, mechanistic, and functional evidence, we provide an overview of the most effective enzyme modifications, discuss their underlying rationales and implications for IVT optimization. Finally, we identify the remaining challenges in engineering T7~RNAP variants for the development of robust, high-purity, and scalable IVT systems for therapeutic RNA manufacturing.


\section{Structural and functional properties of T7~RNAP}

\subsection{T7~RNAP structure}

T7~RNAP is a single-subunit enzyme of 883 amino acids with a molecular weight of approximately 98~kDa. Its three-dimensional structure is organized into an N-terminal domain (NTD) and a C-terminal polymerization domain (CTD), which function in a coordinated manner to perform promoter recognition, catalysis, and processive elongation. The X-ray structures of T7 RNAP in the transcription initiation and elongation conformations are shown in Figure~\ref{fig:rnap_structures} and Supplementary Figure~S1.

The NTD (residues~1--312) plays a central role in promoter recognition and escape, as well as in mediating interactions with DNA and nascent RNA. It also contributes to CTD stabilization and undergoes extensive conformational rearrangements during the transition from transcription initiation to elongation. It is composed of several specialized subdomains: the C-helix (residues~28--71) facilitates active-site enlargement during transcription initiation \cite{dousis2023} and also promotes the transition from transcription initiation to elongation; 
the promoter binding subdomain (residues~72--150 and~191--250) makes sequence-nonspecific interactions with the DNA promoter and promotes DNA duplex opening through its intercalating loop \cite{sousa2003, tang2009};
the H subdomain (residues~151--190) mainly  contributes to the formation of the RNA exit channel and helps stabilize the nascent RNA transcript during elongation \cite{tang2009, yin2002}; 
and the flexible C-linker (residues~251--296), which connects the NTD and CTD,  enables the domain rearrangements associated with the initiation-elongation transition \cite{dousis2023}. 

The CTD (residues~313--883) adopts a right‑hand–like architecture composed of ``thumb'', ``palm'', and ``fingers'' subdomains that form a deep, positively charged nucleic‑acid‑binding cleft \cite{nair2024, sousa2003}. 
The palm subdomain (residues~411--448, 532--540, and~788--838) forms the catalytic core and contains catalytic and metal-binding residues essential for the catalytic mechanism and phosphodiester bond formation \cite{sousa2003}.
The fingers subdomain (residues~541--738 and~771--787) is composed of five helices, one of which, the O-helix, is highly conserved among RNA and DNA polymerases and plays a crucial role in the catalysis of the polymerization reaction as it binds incoming ribonucleoside triphosphates and contributes to substrate selectivity and fidelity \cite{steitz2009, temiakov2004, wu2017}.
The thumb subdomain (residues~330--410) stabilizes the RNA:DNA hybrid, ensuring elongation complex stability \cite{sousa2003, durniak2008}. 
Additional CTD elements, including the specificity loop (residues~739--770), an extra helical bundle (residues~449--531), and a C-terminal loop (residues~839--883), further contribute to promoter specificity, nucleotide binding, and regulatory interactions \cite{sousa2003, nair2024}.

\subsection{T7~RNAP function}

Transcription by T7~RNAP proceeds through three distinct phases: initiation, elongation, and termination, each accompanied by substantial conformational rearrangements, particularly within the NTD (see Figure~\ref{fig:rnap_structures} and Supplementary Figure~S1).

During initiation, the promoter binding subdomain of the NTD and the specificity loop of the CTD selectively recognize and bind the T7~promoter, followed by local DNA melting mediated by the NTD intercalating loop (residues~232--242).
Early initiation involves the formation of a highly unstable transcription complex. As the nascent RNA extends to~3--5 nucleotides, the growing RNA:DNA hybrid exerts steric pressure on the NTD, triggering its rotation away from the CTD and leading to formation of a late initiation complex (IC) that still retains promoter contacts \cite{nair2024, sousa2003}.
Throughout all the initiation phase, abortive cycling is frequent, resulting in the release of short abortive transcripts (SATs) due to destabilization of the RNA:DNA hybrid, consistent with the push‑back model \cite{cheetham1999, durniak2008, ramirez2012}. 

Once the nascent RNA molecule reaches approximately~8--12 nucleotides in length, a major structural transition occurs, leading to the conversion of the IC into a stable elongation complex (EC). This conformational rearrangement primarily involves the NTD, in particular the C-helix and C-linker subdomains which transition from helix-turn-helix to simple helical conformations. Consequently, promoter-binding interactions are dismantled, the nucleic-acid-binding cleft expands to accommodate a~7--8~bp RNA:DNA hybrid, and the RNA exit channel is formed. These structural changes give rise to a highly processive, sequence-independent elongation phase in which the polymerase slides along the DNA template while efficiently synthesizing RNA transcripts over several kilobases \cite{nair2024, steitz2009, yin2002}. 

The final phase is termination, during which the EC is destabilized through the loss of key interactions. Consequently, the nascent RNA transcript is released and T7~RNAP dissociates from the template. This phase is intrinsically inefficient, particularly \textit{in vitro}, due to the high stability of the EC, which favors continuous RNA elongation over termination \cite{beckert2010, nair2024}. This frequently results in the generation of run-off transcripts, in which the polymerase continues transcription until reaching the end of the DNA template. Moreover, the destabilization of the EC at the template end frequently results in heterogeneous 3'-RNA termini, including transcripts with extra or truncated nucleotides \cite{beckert2010, nair2024}.


\section{By-product generation during \textit{in vitro} transcription}

\begin{figure}
    \centering
    \includegraphics[width=\linewidth]{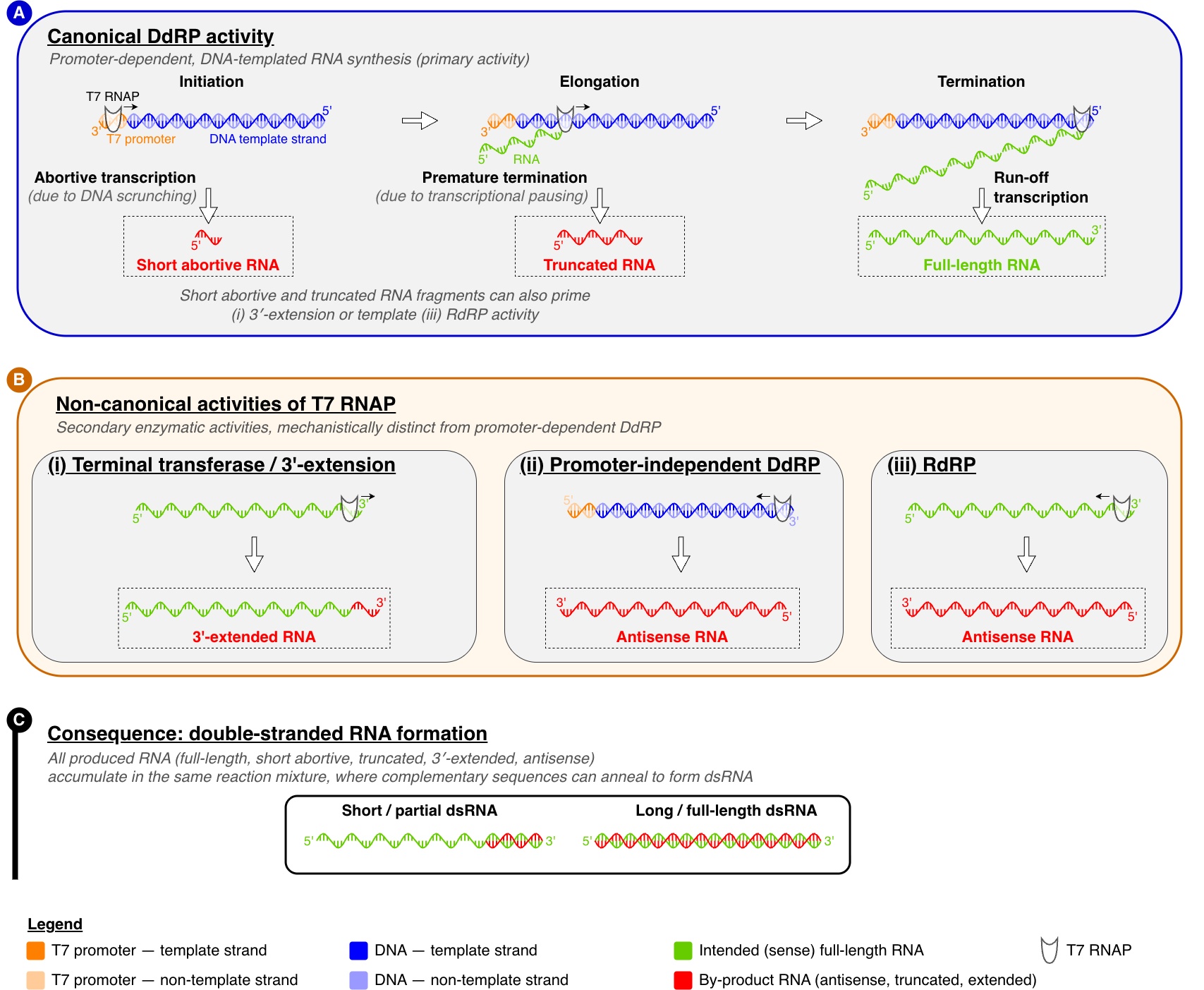}
    \caption{
    \textbf{Canonical and non-canonical activites of T7~RNAP generate RNA by-products during \textit{in vitro} transcription.}
    T7~RNAP catalyzes DNA-dependent RNA polymerization (DdRP) to produce full-length single-stranded RNA from a promoter-containing DNA template, and additionally exhibits several secondary, mechanistically distinct activities that generate RNA impurities.
    \textbf{(A) Canonical DdRP activity.} Promoter-dependent, DNA-templated RNA synthesis proceeds through initiation, elongation, and termination. Disruptions of this canonical pathway generate non-productive RNA species: abortive transcription, driven by DNA scrunching during initiation, releases short abortive RNAs; premature termination, promoted by transcriptional pausing during elongation, releases truncated RNAs; successful run-off transcription yields full-length RNA. Short abortive and truncated RNA fragments can also re-enter the reaction as primers for (i) 3'-extension or templates for (iii) RdRP activity. 
    \textbf{(B) Non-canonical activities of T7~RNAP.} Three secondary enzymatic activities, mechanistically distinct from promoter-dependent DdRP, generate additional RNA by-products: (i) terminal transferase/3'-extension activity adds non-templated or self-templated nucleotides to the 3'-end of RNA transcripts, yielding 3'-extended RNA; (ii) promoter-independent DdRP initiates transcription from DNA termini lacking a canonical T7~promoter, using the DNA as template, to produce antisense RNA; (iii) RNA-dependent RNA polymerase (RdRP) activity uses an RNA molecule as template to synthesize a complementary antisense RNA. 
    \textbf{(C) Consequence: double-stranded RNA (dsRNA) formation.} All RNA species produced in (A) and (B)---full-length, short abortive, truncated, 3'-extended, and antisense RNA---accumulate together in the reaction mixture, where complementary sequences can anneal to form either short/partial dsRNA (e.g., from 3'-extension) or long/full-length dsRNA (e.g., from antisense transcription).
    Intended full-length sense RNA is shown in green; unintended by-product RNA in red; the DNA template and non-template strands are shown in dark and light blue, respectively, and the T7 promoter in dark and light orange.}
    \label{fig:ivt_byproducts}
\end{figure}

Here, we describe the principal by-products generated by T7~RNAP during IVT reactions, namely truncated RNAs and dsRNAs of various lengths. These by-products arise from secondary or non-canonical activities of T7~RNAP, illustrated in Figure~\ref{fig:ivt_byproducts}. 
Indeed, in addition to its primary promoter-dependent DNA-dependent RNA polymerase (DdRP) activity, which initiates transcription from a specific T7~promoter and synthesizes RNA using the DNA template, T7~RNAP also exhibits : (i) 3'-extension activity, whereby the enzyme adds nucleotides to the 3'-end of RNA molecules after transcription; (ii) promoter-independent DdRP activity, initiating RNA synthesis from DNA templates lacking a canonical T7~promoter; and (iii)  RNA-dependent RNA polymerase (RdRP) activity, in which RNA molecules serve as templates for the synthesis of complementary RNA strands. 
Furthermore, abortive transcription and transcriptional pausing events can compromise canonical DdRP activity and contribute to RNA heterogeneity  \cite{wu2020, gholamalipour2018, mu2018, he2024}.
These impurities must be minimized, as they can elicit adverse immunogenic responses in host cells. Consequently, to ensure the safety of RNA therapeutics, regulatory agencies impose stringent requirements on product purity and manufacturing quality control.

\subsection{Production of truncated RNA}

RNA molecules shorter than the intended product can be released during IVT as a result of premature transcription termination by T7~RNAP \cite{he2024, nowak2026}. During the initiation phase, SATs, typically fewer than 15~nucleotides in length, are frequently released as a consequence of DNA scrunching, a process in which RNA polymerase remains anchored to the promoter while pulling downstream DNA into the active site, generating mechanical stress that can hinder the transition from initiation to elongation \cite{sousa2003, nair2024, steitz2009}.
Interestingly, the flexibility and conformational properties of the template strand downstream of the promoter have recently been shown to directly influence the stability of the initiation complex by T7~RNAP \cite{zernia2025}.
Following abortive transcription, T7~RNAP can either remain associated with the promoter and initiate a new round of RNA synthesis (recycling), or dissociate from the template before reinitiating transcription in a subsequent binding event (exchange) \cite{koh2018correlating}. Only about~56\% of transcription initiation events by T7~RNAP proceed to productive elongation, while the remaining~44\% terminate prematurely, generating abortive transcription by-products \cite{lenk2024}.

Although the majority of truncated RNAs generated during IVT are SATs, longer truncated transcripts can also be released during the elongation phase. As an example, only approximately~90\% of polymerases that successfully escape the abortive cycle and proceed into the elongation phase produce full-length transcripts when transcribing a~10,000-nucleotide template \cite{martin1988}. Despite the high intrinsic processivity of T7~RNAP, transient transcriptional pauses are frequently observed during elongation. These pauses are sequence-dependent, and their mean duration varies widely, ranging from subsecond intervals to several tens of seconds \cite{zernia2025}. However, pausing is not obligatory, as only a fraction of polymerases pause at specific sites. Although transient pausing generally only reduces transcriptional efficiency, in some cases it can also promote premature termination and the release of truncated RNA products \cite{zernia2025, he2024, he2024b}.

In addition to T7~RNAP-related mechanisms, fragmented RNAs can also arise during IVT through hydrolytic degradation of full-length RNA transcripts in the reaction buffer \cite{he2024, nowak2026, he2024b}. This hydrolysis proceeds via nucleophilic attack by the 2'-hydroxyl group on the adjacent phosphate ester bond, cleaving the phosphodiester backbone  \cite{abouhaidar1999}. The reaction is favored at pH above~7.5 and at temperatures exceeding 37°C \cite{abouhaidar1999, miller2024}, and is further promoted by divalent cations such as $\text{Mg}^{2+}$ \cite{he2024, abouhaidar1999, miller2024}. Since $\text{Mg}^{2+}$ is a required cofactor of T7~RNAP, its presence in IVT reactions is unavoidable, but its concentration can still be tuned to modulate the hydrolysis rate. Local RNA secondary structure and sequence also influence hydrolysis rate, as base‑paired and conformationally constrained regions resist in‑line cleavage more than effectively flexible single‑stranded regions \cite{abouhaidar1999, wayment2021}.

The generation of RNA fragments during IVT inevitably reduces the overall reaction yield and impairs translation efficiency, because truncated transcripts are incapable of encoding full-length open reading frames. Moreover, these fragments can indirectly promote dsRNA formation by serving as primers for 3'-extension \cite{gholamalipour2018, zernia2025} or as templates for antisense transcription \cite{nowak2026}. Truncated RNAs are also intrinsically immunogenic, not only because they facilitate dsRNA formation, but also because they share structural and chemical features with pathogenic nucleic acids, such as the absence of a 3'-poly(A) tail, that are recognized by innate immune receptors \cite{roers2016}.

\subsection{Production of double-stranded RNA}

The most prevalent impurity generated during IVT reactions is dsRNA, which arises primarily from promoter-independent transcriptional activities of T7~RNAP. Both short (``partial'') dsRNA species and long (``full-length'') dsRNA molecules, the latter being comparable in length to the intended single-stranded RNA (ssRNA) transcripts, can be produced by T7~RNAP \cite{yu2024}.
Depending on the reaction conditions, dsRNA by-products can account up to approximately one-third to one-half of total RNA produced during IVT \cite{mu2018}.
Both classes of dsRNA act as a global cellular danger signal, where increasing amounts sequentially trigger innate immune activation, translation shutdown, and ultimately cell death. These consequences make dsRNA a major impurity of IVT that can severely compromise mRNA therapeutic efficacy \cite{cieslicka2026, liu2025}.

The molecular mechanisms underlying dsRNA formation during IVT are not yet fully resolved. Nevertheless, two major classes of IVT by-products have been identified as principal contributors to dsRNA generation \cite{wu2020, he2024, jo2025}: 

\begin{itemize}
	
	\item 3'-extended by-products are generated from run-off transcripts whose sequences are further extended at their 3'-end by T7~RNAP \cite{cazenave1994rna, gholamalipour2018, tang2025}. The newly added 3'-extension can anneal to complementary regions within the same RNA molecule (\textit{cis} mechanism) or with a different RNA molecule (\textit{trans} mechanism), thereby promoting the formation of dsRNA structures \cite{gholamalipour2018, he2024, wu2020}. 
    Experimental evidence suggests that the 3'-extension proceeds via a distributive mechanism, in which the polymerase adds one or a few nucleotides before releasing the extended RNA, which can subsequently rebind the enzyme for further extension. This behavior leads to substantial length heterogeneity among 3'-extended by-products \cite{gholamalipour2018}.

    There is currently no consensus regarding the mechanism responsible for the initial addition of nucleotides at the 3'-end of run-off transcripts. Some studies have proposed that these additions arise from the RdRP activity of T7~RNAP, whereby an RNA molecule serves as the template for 3'-extension \cite{gholamalipour2018, cazenave1994rna}. According to this hypothesis, 3'-extension is initiated by the formation of two or three base pairs, either in \textit{cis} or in \textit{trans}, with the resulting short duplexes serving as primers for RNA-templated 3'-extension \cite{gholamalipour2018, cazenave1994rna}.
    
    Alternative explanations propose that the initial nucleotide additions at the RNA 3'-end arise through a template-independent mechanism mediated by T7~RNAP, followed by further elongation through its RdRP activity \cite{tang2025}. Indeed, T7~RNAP possesses a terminal transferase activity analogous to that of Taq DNA polymerase, which consistently appends a non-templated adenosine residue to the 3'-ends of nascent DNA strands \cite{brownstein1996}. In the case of T7~RNAP, approximately~25\% and~32\% of transcripts have been reported to contain one or two additional cytosines, respectively, at their 3'-ends \cite{tang2025}. The high prevalence of cytosine additions challenges a model relying solely on RdRP activity for 3'-extension, although the sequence of the DNA template terminus likely influences which nucleotides are added to the RNA 3'-end \cite{tang2025}.

	\item Antisense by-products can be fully complementary to the desired RNA, enabling the formation of long dsRNA upon hybridization. A major and broadly supported source of antisense transcripts in T7~RNAP IVT is promoter-independent transcription initiated at promoter-less termini of linear DNA templates, generating antisense RNA that anneals with the sense product \cite{mu2018, lee2026}. 
	The structural determinants that enable T7~RNAP to initiate productive transcription from DNA termini remain incompletely understood. Mapping of antisense 5'-ends indicates that initiation frequently occurs at the penultimate position of the promoter-less DNA terminus \cite{yu2024}. The enzyme would start the transcription directly in an elongation-like conformation, bypassing standard promoter recognition \cite{mu2018}.
    Importantly, DNA-terminus-initiated antisense synthesis depends on sequence and structural features of the DNA end \cite{lee2026, mu2018, yu2024}. In particular, terminal guanine or cytosine residues 
    enhance the generation of full-length dsRNA through this pathway \cite{yu2024}. Conversely, template-encoded poly(A) tails (as is the case for most mRNA therapeutic applications) partially suppress antisense by-product formation \cite{wu2020, lee2026}.
    
    In addition to DNA-end initiation, T7~RNAP has been reported to catalyze RNA-templated RNA synthesis in highly specific systems, generating complementary strands efficiently and accurately. However, this activity appears strongly constrained by RNA sequence and structure contexts and has limited generality \cite{konarska1989}. Overall, available evidence supports DNA-end-initiated, promoter-independent transcription as the dominant source of antisense-derived full-length dsRNA during T7~RNAP IVT \cite{mu2018, lee2026, yu2024}.
    	
\end{itemize}

Promoter-independent transcriptional activities of T7~RNAP are intrinsically inefficient relative to canonical promoter-driven initiation. However, they become increasingly prevalent under high-yield batch IVT conditions. As run-off RNA accumulates in the reaction mixture, the likelihood of its rebinding to T7~RNAP increases, thereby promoting the formation of 3'-extended by-products or antisense transcripts via RNA-templated mechanisms \cite{gholamalipour2018}. Elevated concentrations of T7~RNAP further exacerbate these effects.

At present, there is no consensus regarding the dominant pathway responsible for dsRNA production during IVT. The relative contribution of the mechanisms described above likely depends on reaction conditions and template design. In the presence of DNA 3'-end sequences that promote polymerase switching toward the non-template strand, dsRNA production is expected to be dominated by antisense transcription from DNA templates. In contrast, in the absence of such sequences (for example, when the DNA template encodes a poly(A) tail at its 3'-end), RNA-templated by-products are likely the primary source of dsRNA formation \cite{wu2020}.


\section{Process engineering strategies to reduce T7~RNAP by-product formation}

By-products generated during IVT can be removed after transcription using purification techniques such as reverse-phase high-pressure liquid chromatography (RP-HPLC) \cite{kariko2011, weissman2012} or cellulose chromatography \cite{baiersdorfer2019}. However, post-transcriptional removal of IVT by-products is challenging because these impurities share similar sizes and physicochemical properties with the intended RNA products. Consequently, downstream purification strategies typically increase manufacturing costs and reduce overall yield, making prevention of by-product formation during IVT a more efficient and economical approach.

Before focusing on enzyme engineering strategies, it is useful to briefly summarize alternative approaches that suppress by-product formation during IVT, highlighting both their utility and inherent limitations. A variety of strategies based on reaction conditions or reagent selection have demonstrated efficacy in reducing by-product formation, including:
\begin{itemize}
    
    \item The use of modified nucleotides partially suppresses the formation of DNA-templated antisense by-products \cite{mu2018, baiersdorfer2019, nance2021} as well as 3'-extended by-products \cite{he2024}. The molecular mechanisms by which modified ribonucleoside triphosphates inhibit dsRNA production remain unclear.    
    Incorporation of modified nucleotides such as $N1$-methylpseudouridine during IVT can reduce both the amount and immunostimulatory potency of dsRNA by limiting dsRNA formation, decreasing immune recognition of RNA secondary structures, and weakening interactions between immune sensors and ssRNA \cite{nelson2020, nance2021, mu2021}. In parallel, modified uridines enhance translation efficiency by improving ribosome loading and elongation, and stabilizing beneficial RNA secondary structures. Together, these effects decouple immune sensing from protein production, although sequence-specific contexts may occasionally compromise therapeutic performance even with modified nucleotides \cite{nance2021}.

    \item Reducing the $\text{Mg}^{2+}$ concentration decreases the formation of antisense RNA transcribed from the non-template DNA strand, likely because promoter-independent transcription is more sensitive to suboptimal reaction conditions than promoter-driven transcription \cite{mu2018}. However, because Mg$^{2+}$ is an essential cofactor for T7~RNAP catalysis, lowering its concentration also reduces overall transcriptional yield \cite{wu2020}. $\text{Mg}^{2+}$ ions also have a significant impact on RNA stability, playing a dual role: they promote RNA hydrolysis, but also induce RNA folding that protects vulnerable phosphodiester bonds from hydrolysis \cite{guth2023}. 
    Given the importance of $\text{Mg}^{2+}$ for both the IVT reaction and the quality of its products, carefully balancing its concentration in the reaction buffer is essential to find a trade-off that maximizes its positive effects (IVT yield and RNA folding), while minimizing its negative ones (dsRNA generation and RNA hydrolysis).

    \item The addition of chaotropic agents, such as urea and formamide, creates a mildly denaturing IVT environment that limits undesirable inter- and intramolecular base-pairing of RNA molecules, thereby reducing dsRNA formation arising from both 3'-extension and sense–antisense hybridization mechanisms \cite{piao2022, he2024, francis2024}, while only minimally affecting overall RNA yield. The presence of urea during IVT has also been reported to reduce the generation of truncated transcripts \cite{francis2024}.
    
    \item High-salt conditions (e.g., elevated NaCl concentrations) destabilize electrostatic interactions between positively charged residues on the  RNAP surface and the negatively charged phosphate backbone of DNA or RNA. Under such conditions, most protein-nucleic acid interactions, including initial binding of T7~RNAP to its promoter and rebinding of run-off RNA, are weakened. This reduces the production of 3'-extended by-products but also inhibits the intrinsic transcriptional activity of T7~RNAP \cite{cavac2021, malagodapathiranage2024}.
    
    \item DNA templates encoding a poly(A) tail partially suppress the formation of DNA-templated antisense RNAs, likely because the corresponding poly(dT) 
    tract disfavors promoter-independent transcription from the non-template DNA strand \cite{wu2020, lee2026}.

    \item Modified promoter sequences that weaken upstream promoter interactions with T7~RNAP reduce the production of abortive transcripts by facilitating promoter release of the growing DNA:RNA hybrid. However, this strategy is also associated with a reduced overall transcriptional yield \cite{tang2014, lenk2024}.

    \item Site-specific nicks introduced into the non-template DNA strand, either within the promoter region \cite{malagodapathiranage2023} or near the 3'-end of the template \cite{lee2026}, can also reduce antisense RNA formation. When introduced in the promoter region, such nicks may additionally reduce abortive transcript production under certain conditions, but this effect is often accompanied by a substantial decrease in overall RNA yield \cite{jiang2001}.

    \item Addition of a DNA oligonucleotide complementary to the terminal nucleotides at the 3'-end of run-off transcripts suppresses 3'-extension by protecting this terminus. However, this approach introduces an additional contaminant into the reaction mixture that must be removed during post-IVT purification \cite{gholamalipour2019}.
    
    \item Purification of linearized plasmid DNA templates by RP-HPLC reduces by-product formation by eliminating DNA impurities that can serve as unintended templates for transcription of RNA fragments complementary to run-off transcripts \cite{martinez2023}. However, this means having one additional and costly step for RNA production.
        
\end{itemize}

Despite a positive impact on reduction of ultimate dsRNA content, these approaches often compromise RNA yield and increase process complexity or introduce additional contaminants. An alternative strategy, addressed in depth in the next section, consists of engineering T7~RNAP itself to reduce by-product formation arising from its intrinsic non-canonical activities.


\section{Protein engineering strategies to reduce T7~RNAP by-product formation}

This section reviews the various T7~RNAP engineering strategies that have been developed to reduce IVT by-product formation. The enzyme modifications and the mechanistic basis for their effects are discussed in detail in the following subsections. A schematic overview of these strategies is provided in Table~\ref{tab:T7RNAP_strategies}.

\begin{figure}[p]
    \centering
    \includegraphics[width=0.9\linewidth]{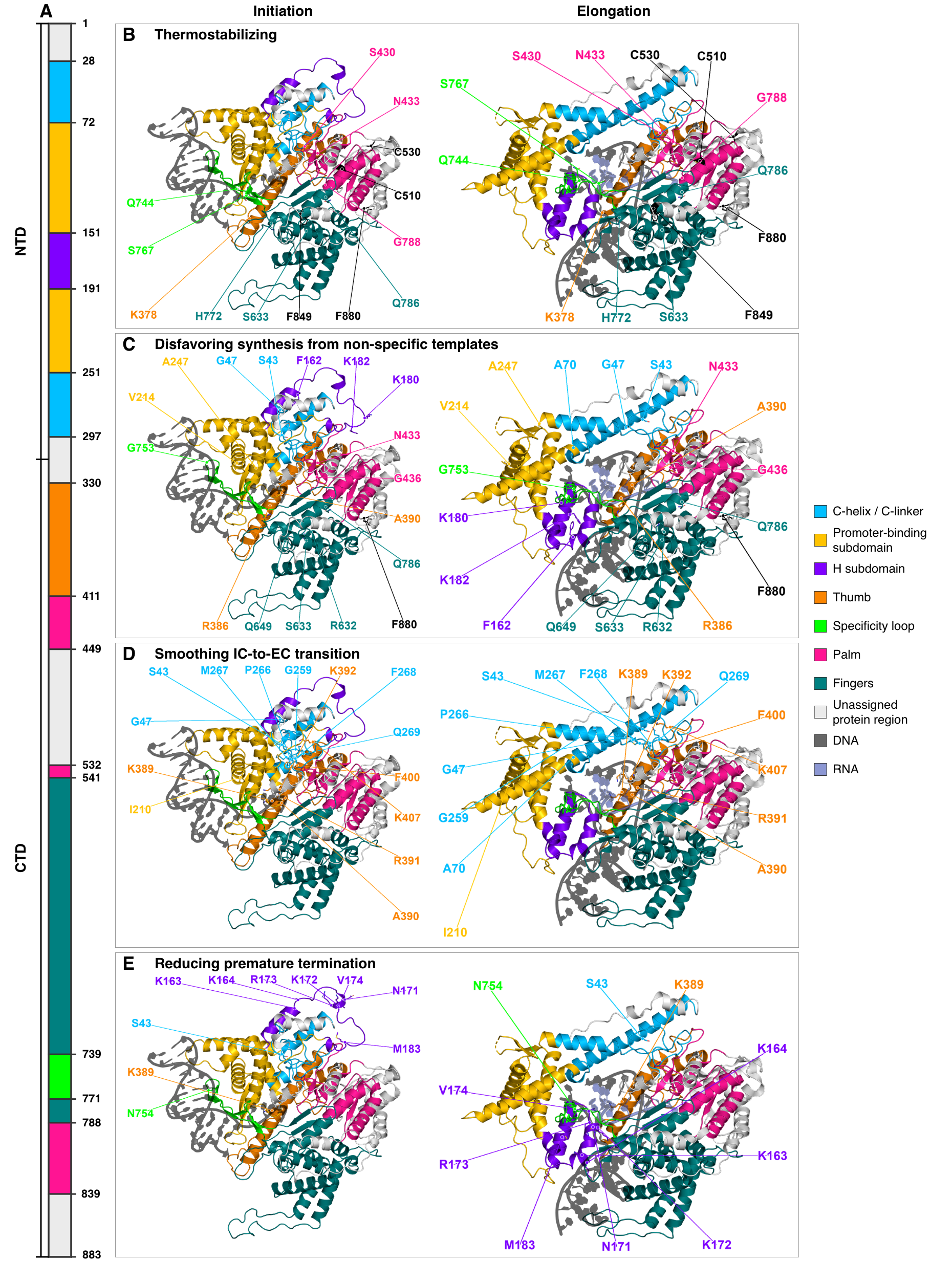}
    \caption{
    \textbf{Structural mapping of engineered T7~RNAP mutations onto the initiation and elongation complexes.}
    (A) Sequential domain organization of T7~RNAP (883 residues), shown from N- to C-terminus (top to bottom). The N-terminal  (NTD, residues 1-313) and C-terminal  (CTD, residues 314-883) domains are indicated  at the left with tick marks denoting subdomain boundaries. Named subdomains are colored according to the key and consistently across all panels (A–E); unassigned regions are shown in light grey.
    (B-E) Cartoon representation of the initiation complex (IC; PDB 1CEZ, left) and elongation complex (EC; PDB 1MSW, right), with DNA and RNA (EC only). Residues of each mutation category are shown as sticks and labeled according to their subdomain color, or in black if outside the named subdomains. Residue A70, unresolved in the 1CEZ structure, is not shown in the IC structures of panels C and D.}
    \label{fig:rnap_structures}
\end{figure}

\subsection{Thermostable T7~RNAP variants} \label{subsec:mutants_thermostability}

\subsubsection{Advantages and limitations of high-temperature IVT}

Increasing the IVT temperature is a robust process-level strategy to improve RNA purity, primarily by suppressing RNA-templated non-canonical reactions. In particular, high-temperature transcription reduces the formation of 3'-extended by-products, but does not appear to significantly reduce antisense products initiated from DNA termini under the conditions tested \cite{wu2020}.
From a biophysical perspective, elevated temperatures reduce dsRNA by-product formation through several mechanisms: (i)~decreasing the affinity of run-off transcripts for T7~RNAP, thereby limiting transcript rebinding and subsequent RNA-templated self-priming events; (ii)~destabilizing RNA secondary structures of the nascent transcripts, thereby reducing conformations that promote 3'-extension; and (iii)~reducing the stability of the short RNA:RNA duplexes required to initiate extension reactions \cite{wu2020}. Additionally, higher temperatures may facilitate conformational transitions within T7~RNAP, potentially improving promoter clearance and reducing aberrant transcription products \cite{wu2020}. 

These  mechanisms primarily explain the reduction of RNA-templated 3'-extension but do not explicitly address antisense RNA synthesis initiated from sense RNA templates, an additional source of dsRNA \cite{konarska1989}. If elevated temperature indeed decreases the affinity of T7~RNAP for RNA templates, this effect would be expected to suppress not only 3'-extension but also RNA-templated antisense synthesis. In contrast, antisense transcription initiated directly from DNA termini would likely be less affected, as it does not rely on RNA serving as the template. In therapeutic mRNA production, where template-encoded poly(A) tails already partially suppress promoter-independent transcription from DNA ends, high-temperature IVT may therefore further reduce dsRNA formation by simultaneously limiting RNA-templated by-products \cite{wu2020}. The same operating regime can also benefit circular RNA manufacturing, where elevated temperature has been identified as a key determinant for achieving high cyclization ratios \cite{he2024c}.

However, high temperatures accelerate RNA degradation by increasing its hydrolysis rate, thereby limiting the operational window. Moreover, wild-type T7~RNAP is stable and active up to approximately~41°C, but becomes progressively less active at higher temperatures. Finally, even at~50°C, transcripts containing one or two non-templated nucleotides at their 3'-termini can still be observed, indicating that the terminal transferase-like activity of T7~RNAP is not fully eliminated by temperature alone \cite{wu2020, tang2025}.

These considerations support the use of thermostable polymerase variants to enable IVT reaction in the temperature range where dsRNA suppression becomes significant (reported from approximately~48°C, around~7°C higher than the stability range of wild-type T7~RNAP), while maintaining sufficient RNA integrity and catalytic activity \cite{wu2020, boulain2013}.

\subsubsection{Engineering thermostable T7~RNAP variants for improved IVT purity}

Before dsRNA emerged as a major concern in IVT quality control, numerous studies had already explored strategies to increase the thermal resistance of T7~RNAP \cite{liao2003, sugiyama2009, oe2011, boulain2013, meyer2015}. While these studies generally did not quantify dsRNA formation, they identified thermostabilizing substitutions that can be repurposed for high-temperature IVT, where reduced dsRNA is a downstream consequence of the operating conditions rather than a direct effect of mutations. Building on this concept, more recent studies have specifically engineered thermostable T7~RNAP variants to directly reduce dsRNA formation during high-temperature IVT \cite{he2024c, wu2020, qin2026, jiang2026deep}.

Several strategies have been employed to improve the thermal stability and activity of T7~RNAP, including random mutagenesis coupled with functional screening, iterative directed evolution, structure-based computational design and, more recently, deep learning-guided protein engineering combined with large-scale experimental mutagenesis. These approaches have identified numerous thermostabilizing substitutions distributed throughout both the NTD and CTD, with several recurrent mutational hotspots emerging across independent studies. Here, we review the best-characterized variants, illustrated in Figure~\ref{fig:rnap_structures}B, while a comprehensive overview of all engineering efforts reported to date is provided in Supplementary Section S2. 

Wild-type T7~RNAP exhibits a melting temperature ($T_m$) in the range of~40--44°C, depending on the experimental conditions, but its transcriptional activity decreases rapidly as the reaction temperature increases above~37°C, limiting its applicability in high-temperature IVT reactions. Indeed, wild-type T7~RNAP exhibits approximately tenfold lower activity at~45°C than at~37°C \cite{he2024}.  

The earliest studies to identify single-site T7~RNAP variants with enhanced thermal stability employed random mutagenesis \cite{liao2003, sugiyama2009, oe2011, boulain2013}.
These studies identified several beneficial mutations, with S430P, N433T, S633P, F849I, and F880Y  consistently conferring improved thermostability \cite{liao2003, sugiyama2009, boulain2013, meyer2015}. These mutations are located within the CTD and stabilize T7~RNAP through distinct mechanisms. S430P and N433T strengthen interactions between the CTD and NTD, thereby enhancing interdomain stability. In contrast, F849I and F880Y primarily stabilize intradomain interactions within the CTD \cite{meyer2015}. The S633P mutation is located in the O-helix of the fingers subdomain, adjacent to the active site, where it may introduce a kink in the helix that stabilizes the local conformation \cite{liao2003}.

Although several of the many single-site mutations tested enhanced enzyme thermal stability, the gains were limited. The variants exhibited only modest increases in half-life at elevated temperatures and, importantly, retained substantially lower catalytic activity than the wild-type enzyme at physiological temperature. This reflects the well-known activity–stability trade-off, whereby stabilizing substitutions often increase structural rigidity at the expense of the conformational plasticity required for efficient catalysis \cite{hou2023enzyme}.
This is a particularly important consideration for T7~RNAP, which has relatively low intrinsic stability and undergoes extensive conformational rearrangements during the transcription cycle \cite{boulain2013}. Indeed, previous studies have reported a clear anticorrelation between thermal stability and enzymatic activity for T7~RNAP. For example, the thermal stability and activity descriptors of 54 single-point mutants were found to be negatively correlated (Pearson's $r\approx-0.6$) \cite{jiang2026deep}, highlighting the inherent difficulty of simultaneously optimizing both properties. On average, mutations that increased thermal stability reduced enzymatic activity by approximately~30\%, with the most thermostabilizing variants often being almost completely inactive.

To achieve greater improvements in thermostability while maintaining enzymatic activity, early engineering efforts largely relied on combining the most promising single-site mutations identified through experimental screening. One of the first successful examples is the V7abcd variant, which carries the seven substitutions S430P, C510R, Q744R, S767G, Q786L, F849I, and F880Y. Originally developed in \cite{boulain2013}, V7abcd retained approximately~90\% of the wild-type specific activity at~41°C and exhibited an approximately~10°C increase in half-life temperature after 10~min incubation. Similarly, the M5 variant reported in \cite{meyer2015}, carrying the substitutions S430P, N433T, S633P, F849I, and F880Y, retained transcriptional activity comparable to that of the wild-type enzyme at~37°C, while exhibiting markedly improved activity at~50°C. 

A number of additional beneficial single-site mutations have subsequently been identified in different studies \cite{meyer2015, ong2021, he2024c, baumer2026, qin2026, jiang2026deep}, with most subsequent engineering efforts aimed at combining them with previously developed variants through iterative rounds of mutagenesis.
For example, the M5 variant was improved by introducing the G788A substitution in \cite{he2024c}, and the K378R substitution in \cite{baumer2026}.
Similarly, the M5 and V7abcd variants were combined and further engineered by introducing additional substitutions, resulting in the variant $\mathrm{T7}^{\mathrm{T+}}$ containing 30 mutations, which has a half-inactivation temperature of 54.9°C \cite{baumer2026}. However, this variant retained only~59\% of the wild-type activity at~37°C.
The most thermostable T7~RNAP reported to date ($T_m=$ 55.7°C), composed of 14 mutations, exhibited a 60-fold increase of activity at 52°C compared to the wild-type \cite{jiang2026deep}.

The predominant strategy adopted in these studies was to combine thermostabilizing mutations with compensatory substitutions that, while not increasing thermostability on their own, mitigated the activity loss caused by other stabilizing mutations. Notably, Q744R and H772R have been reported to restore or preserve polymerase activity when combined with otherwise deleterious thermostabilizing substitutions \cite{boulain2013}.

More recently, different computational approaches have guided the engineering of T7~RNAP, thereby accelerating the identification of beneficial variants and, in particular, the selection of variant combinations. In \cite{he2024c}, multiple computational methods \cite{sumbalova2018hotspot, dehouck2011popmusic, capriotti2005mutant2} were used to identify stability hotspots, including C530S and G788A. These mutations were introduced into the M5 variant background and experimentally evaluated, with the M5/G788A variant exhibiting stable activity at~45°C.
Using a similar approach, \cite{baumer2026} employed the structure-based protein design algorithm PROSS~2.0 \cite{weinstein2021pross} to identify beneficial mutations that were subsequently incorporated into the M5 and V7abcd variants. In \cite{jiang2026deep}, a protein language model was first used to predict promising single-site variants, which were experimentally evaluated in an initial screening round. The models were then fine-tuned using the experimental data and iteratively applied to guide the selection of beneficial variant combinations.

While T7~RNAP engineering strategies have successfully increased enzyme thermostability, sometimes at the expense of catalytic activity, the complexity of the molecular interactions underlying protein thermal stability \cite{pucci2017physical} makes it difficult to rationalize the effects of individual mutations and their combinations. Indeed, the identification of stabilizing substitutions in nearly all regions of both domains of T7~RNAP further indicates that thermostability results from a distributed interaction network rather than from isolated structural elements.

Only some of these enzyme engineering studies have experimentally evaluated the reduction of dsRNA formation as an indirect consequence of improved T7~RNAP thermostability, which enables IVT to be performed at higher temperatures. The reported results consistently show substantial reductions in dsRNA by-product formation under high-temperature IVT conditions, confirming thermostabilization as a good strategy for improving IVT purity. For example, the M5/G788A variant enabled IVT at~50°C, maintaining RNA yield and reducing dsRNA by-product formation to less than~20\% of the level observed with the wild-type enzyme during IVT at~37°C \cite{he2024c}.
Similarly, two optimized T7~RNAP variants containing 9 and 10 mutations, constructed in \cite{jiang2026deep}, reduced dsRNA by-product formation during IVT at~50°C to about~30\% of the level produced by the wild-type enzyme.

\subsection{T7~RNAP variants disfavoring synthesis from non-specific templates} \label{subsec:mutants_affinity}

A major fraction of dsRNA produced during IVT originates from non-canonical template usage by T7~RNAP, including RNA-templated extension/synthesis and promoter-independent initiation from DNA termini. A second engineering strategy to achieve high-purity IVT aims to bias the enzyme toward productive, promoter-driven, DNA-templated transcription while disfavoring interactions with non-specific substrates (including run-off RNAs, abortive or truncated RNAs, and promoter-less DNA ends). The proposed mutations of this category are shown mapped onto the T7~RNAP structures in Figure~\ref{fig:rnap_structures}C.

\paragraph{Tuning initiation determinants to suppress non-canonical starts.}
To achieve this, a first class of target residues comprises those that directly mediate initiation-related interactions and are key determinants of promoter-driven transcription initiation. In \cite{he2024}, the authors investigated conserved residues in the specificity loop (R746, G753, T763), which is responsible for sequence-specific recognition and binding of the T7~promoter during transcription initiation, hypothesizing that these positions could be engineered to reduce promoter-independent transcription without compromising canonical promoter recognition. They identified G753A as a strong dsRNA-reducing mutant (approximately $82\%$ reduction relative to wild-type), with other substitutions at the same position (G753N, G753C, G753Q, G753W) also reducing dsRNA, albeit to a lesser extent. Mechanistically, G753A reduced antisense transcription initiated from promoter-less DNA ends as well as antisense formation using RNA as a template, while only partially reducing 3'-extension. This reduction came with a trade-off: the G753A variant exhibited increased RNA fragmentation, presumably due to its reduced affinity for the DNA template \cite{he2024}.

\paragraph{Indirect modulation of substrate recognition and catalysis.}
Another class of target residues comprises those located near the protein–nucleic acid interface that, while not directly contacting the substrate, may reduce nonspecific nucleic acid binding or allosterically bias substrate recognition toward DNA rather than RNA. According to \cite{yu2024}, increasing steric bulk at G47 in the C-helix subdomain, particularly through the G47W substitution, markedly reduces full-length dsRNA formation while only modestly weakening canonical promoter binding; moreover, mRNA transcribed by this variant exhibited higher expression and lower immunogenicity than mRNA generated by the wild-type enzyme \cite{yu2024}. 
Similarly, aromatic substitutions were identified in the thumb subdomain (R386F/Y/W) that strongly reduce dsRNA formation \cite{matsumoto2025}. While the thumb subdomain participates in interactions with the RNA:DNA hybrid in the EC, these substitutions were proposed to reduce non-productive RNA:RNA or RNA:DNA interactions involved in template re-engagement and aberrant initiation rather than to disrupt productive elongation. Additional nearby NTD mutations (E48A, R50A, F51A, R52A, P72A) were suggested as potential synergistic contributors of R386 aromatic substitutions \cite{matsumoto2025}.

Additional variants located in the NTD were identified through directed evolution coupled with semi-rational design and were experimentally verified to substantially reduce dsRNA formation \cite{tang2025}. In particular, a triple mutant A70Q/F162S/K182E reduced dsRNA content to~1.80\% of that observed for the wild-type enzyme. Other variants including the mutations K180E, V214A, and A247T also exhibited reduced dsRNA content to a lower extent.
Additional analyses suggest that these mutations decrease binding affinity for ssRNA, enhance binding to DNA templates, while also significantly reducing the terminal transferase activity of the polymerase \cite{tang2025}. Collectively, these mutations markedly reduce dsRNA generation during IVT, likely through a combination of mechanisms that remains difficult to rationalize from a structural perspective, as the mutated residues do not directly contact either the nucleic acid substrate or the catalytic site.

Finally, high-throughput screening also identified substitutions in catalytically coupled helices within the active-site region that reduce dsRNA formation without compromising RNA yield. Indeed, six variants that reduce dsRNA levels by more than~80\% relative to the wild-type enzyme while maintaining comparable transcription yields were reported in \cite{sanjeev2026}: A390N, A390W, N433D, G436A, R632Q, and Q649L.
Notably, the mutations R632N and Q649L increase also co-transcriptional capping efficiency \cite{sanjeev2026}. Mechanistically, residues R632 and Q649, located in the fingers subdomain, participate in initiating nucleotide interactions, substrate selection, and translocation. Substitutions at these positions remodel the active-site environment by increasing the solvent-accessible cavity volume and altering local residue interactions. Although the precise mechanism remains unresolved, these substitutions are proposed to reshape the energetic landscape of the initiation and elongation states, ultimately reducing the propensity for RNA-dependent RNA synthesis \cite{sanjeev2026}.

\paragraph{Substitutions reshaping non-catalytic cavities.} A third class includes substitutions that reshape solvent-inaccessible cavities near the active site, with the potential to disfavor weak or non-productive substrates (e.g., RNA-templated extension) more than canonical promoter-driven transcription.

A prominent mutational target is the C-terminal "foot" of T7~RNAP (residues 880--883, FAFA), which lies adjacent to a small buried cavity near Mg$^{2+}$-coordinating elements of the active site \cite{gardner1997, dousis2023}. Although foot residues are not directly catalytic, their size and hydrophobicity can influence local packing and restrict access to conformations that accommodate non-specific substrates. 
However, increased steric constraints within the cavity often improve product homogeneity at the expense of overall RNA yield \cite{dousis2023}.

More in detail,  mutations at any of the four positions of the C-terminal foot have been reported to markedly reduce initiation efficiency by weakening interactions with both the promoter DNA and the initiating nucleotide \cite{gardner1997}. Among the substitutions tested, the F880A mutant (AAFA) uniquely produced full-length transcripts while abolishing the terminal transferase activity responsible for non-templated 3'-nucleotide addition, thereby generating more homogeneous RNA products. This improvement in product homogeneity was accompanied by an approximately~50\% reduction in initiation rate, with little effect on elongation processivity \cite{gardner1997}.

Extending these results, variants containing one or two additional residues at the C-terminal “foot” (positions 884 and 885) were generated \cite{dousis2023}. An inverse relationship between cavity volume and RNA 3'~homogeneity was observed: smaller cavities increased 3'~homogeneity. Several insertions improved 3'~homogeneity with only a limited penalty on yield, with 884G providing a particularly favorable balance, increasing 3'~homogeneity while maintaining yields comparable to those of the wild type \cite{dousis2023}. Together, these results support the concept that active-site-adjacent packing provides a tunable target that selectively suppresses RNA-templated extension and related dsRNA-producing pathways.

Finally, a growing body of work suggests that solvent-inaccessible, non-catalytic cavities located far from the active site can also function as allosteric nodes that modulate enzyme activity through conformational changes \cite{tokuriki2009}. Computational structural analyses identified eight cavities in the CTD and four in the NTD. Based on this concept, a semi-rational design of T7~RNAP mutants focused on these solvent-inaccessible cavities to reduce dsRNA by-products was performed in \cite{qin2026}. The three mutations S43E, S633P, and Q786M are associated with reduced dsRNA formation. They were associated with thermostabilizing mutations (discussed in Section \ref{subsec:mutants_thermostability}) to construct the variant M30 (S43E\ms S397W\ms S430P\ms S633P\ms Q744R\ms Q786M). Compared with wild-type T7~RNAP, M30 exhibited an approximately tenfold increase in catalytic efficiency at~37°C, markedly enhanced thermostability with an increase in $T_m$ of~8.5°C, and a roughly tenfold reduction in the production of dsRNA by-products \cite{qin2026}.

\subsection{T7~RNAP variants with a smooth initiation-to-elongation transition} \label{subsec:mutants_ictoec}

A third engineering strategy to increase IVT product purity targets the transition from the IC to the EC, based on the hypothesis that facilitating this transition reduces the release of SATs.  Because SATs can act as primers for RNA-templated extension or antisense synthesis, limiting their production is expected to indirectly reduce short dsRNA formation during IVT.
The IC–EC transition in T7~RNAP involves extensive conformational rearrangements of the NTD, most prominently within the C-helix (residues 28–71) and the C-linker (residues 251–296) subdomains \cite{dousis2023, sousa2003}. In the IC, both regions adopt helix-turn-helix conformations that subsequently refold into simple helices in the EC. The residues undergoing loop-to-helix transitions are primarily residues 42–47 in the C-helix and 258–265 in the C-linker \cite{dousis2023}. 
Perturbing the flexibility, stability, or interaction network of these elements can therefore reshape the energetic landscape of promoter clearance and bias the enzyme toward productive elongation. Interestingly, mutations in other regions of T7~RNAP can also influence the IC-to-EC transition, as described below. Reported mutations are shown mapped onto the T7~RNAP structures in Figure~\ref{fig:rnap_structures}D.

\paragraph{Increasing IC plasticity: C-linker flexibility delays the IC–EC transition.}
Early evidence that IC–EC mechanics influence abortive transcription came from random mutagenesis studies targeting the C-linker. Indeed, several substitutions at residue P266 (P266L, P266A, P266S, P266Y) have been shown to increase initial processivity and reduce abortive cycling, with P266L showing the strongest effect \cite{guillerez2005}. 
Subsequent mechanistic analyses showed that P266L does not primarily weaken promoter binding, as initially proposed. Instead, it slightly impairs promoter melting and reduces affinity for initiating nucleotides, while increasing the flexibility of the C-linker \cite{tang2014, ramirez2012}. This added plasticity allows the enzyme to better accommodate the steric stress imposed by the growing RNA:DNA hybrid during initiation, effectively delaying the IC–EC transition by approximately 1–2 nucleotides. As a consequence, short abortive transcripts (4–7~nucleotides) are strongly reduced, whereas longer abortives (9–11~nucleotides) are modestly increased \cite{tang2014}. Thus, P266L does not eliminate abortive cycling but redistributes it toward longer intermediates.

Importantly, suppressing SATs via increased IC plasticity does not necessarily translate into reduced dsRNA formation. Recent work showed that P266 variants can paradoxically increase dsRNA under certain IVT conditions, illustrating that delayed IC–EC transition may prolong initiation states permissive to RNA-templated side reactions \cite{sanjeev2026}.

\paragraph{Promoting the formation of a stable EC.}
In contrast to the above strategy, a second class of mutations promotes earlier formation of a stable EC, thereby shortening the lifetime of initiation states prone to abortive release. Mechanistically, this can be achieved either by smoothing the IC-EC transition or by stabilizing EC-like conformations directly. 

Short insertions (1-5 amino acids, often glycine or alanine), single-residue deletions, or helix-disrupting substitutions introduced throughout the C-linker weaken its $\alpha$-helical rigidity and facilitate promoter escape by lowering the energetic barrier for NTD rotation \cite{martin2015}. This leads to shortening of the lifetime of the abortive IC, promoting earlier formation of a stable EC. Notably, the authors argued that such variants primarily affect promoter interactions and overall transcription yield rather then specifically targeting abortive cycling \cite{martin2015}. However, their study largely focused on transcriptional processivity and yield, and  did not directly evaluate dsRNA formation, leaving open the question of how such variants perform in modern IVT settings optimized for RNA therapeutics.

A more direct EC-stabilization strategy was introduced by \cite{dousis2023}, who engineered substitutions in the IC loops of the C-helix and C-linker. These mutations were computationally identified by selecting those predicted to preferentially stabilize the EC over the IC, based on the computationally predicted difference in mutation-induced free energy changes between the IC and EC conformations. Five substitutions in the C-helix loop (S43L, S43E, S43R, G47A, and G47L) were identified as promoting the earlier adoption of EC-like helical conformations, resulting in experimentally observed improvements in 3'~homogeneity and reduced innate immune activation, without affecting RNA yield \cite{dousis2023}.
Among these variants, G47A emerged as the most promising candidate, improving 3'-end homogeneity by approximately 20\% and reducing the interferon-$\beta$-mediated innate immune response by 30\%.

Additional evidence supporting EC stabilization as a strategy for suppressing dsRNA formation comes from the analysis of substitutions near the RNA exit channel. In \cite{tang2025}, the authors reported that mutation A70Q, located in the C-helix and positioned within the RNA exit channel in the EC, reduces dsRNA production during IVT. The authors proposed that A70Q stabilizes the EC conformation and facilitates the IC–EC transition, while also enhancing binding to dsDNA relative to RNA, thereby reducing the effective contribution of RNA-templated transcription \cite{tang2025}. 

Note that mutations of the three residues S43, G47, and A70 have also been identified for their negative impact on non-specific template binding, as discussed in Section \ref{subsec:mutants_affinity} \cite{qin2026, yu2024}.

\paragraph{Weakening IC-specific contacts to favor elongation.}
A distinct but complementary route to smoothing the IC-EC transition targets weakening of IC-specific interactions that oppose elongation, effectively favoring earlier EC stabilization. 
For example, mutations that weaken interactions between the C-linker and the thumb subdomains, including substitutions near residues 259 and 266–269 in the C-linker and residues 389–392, 400, and 407 in the thumb subdomain, destabilize IC-specific contact networks and bias the enzyme toward elongation-compatible conformations \cite{martin2015}. These variants reduce early RNA release by lowering the energetic cost of IC-to-EC transition.

In another work \cite{he2026},  a single substitution in the promoter binding domain, I210V, was reported to suppress  dsRNA formation. Residue I210 contributes to stabilizing RNAP–promoter interactions during early initiation, and its substitution is proposed to weaken promoter binding and facilitate earlier promoter clearance \cite{he2026}. Unlike the other mutations, I210V does not directly alter regions that undergo conformational changes during the IC-to-EC transition. Instead, it reduces the lifetime of promoter-bound IC, thereby limiting the accumulation of abortive transcripts. 
These studies provide clear evidence that tuning IC stability can substantially reduce dsRNA formation, even in the absence of direct EC stabilization.

\subsection{T7~RNAP variants with reduced premature transcription termination} \label{subsec:mutants_termination}

In addition to dsRNA, IVT reactions frequently contain truncated RNAs arising from premature transcription termination, pausing-associated release, or destabilization of the EC. These truncated products reduce functional yield and can further contribute to dsRNA formation by acting as primers for RNA-templated extension or as templates for antisense synthesis. Engineering strategies that reduce erroneous termination events therefore target a complementary quality attribute, the RNA integrity.

Mechanistically, intrinsic termination by T7~RNAP is typically coupled to EC destabilization through RNA secondary structure formation and/or weakening of the RNA hybrid, often involving the RNA exit channel and nearby RNA-contacting elements. Accordingly, reported mutations that reduce termination or fragmentation cluster into two partially overlapping classes: (i) variants in or near the RNA exit channel that attenuate intrinsic termination, and (ii) variants that stabilize productive elongation by tuning RNA–protein and RNA–DNA interactions, thereby reducing pausing-associated release. Proposed mutations are shown in Figure~\ref{fig:rnap_structures}E.

\paragraph{RNA exit channel variants that attenuate intrinsic termination.}
Early evidence that transcription termination can be tuned came from systematic mutagenesis of the H~subdomain, a region of the NTD that contributes structural elements to the RNA exit channel. In \cite{lyakhov1997}, multiple mutants exhibiting reduced termination efficiency across several intrinsic transcription terminator sequences and a reduced tendency to produce fragmented transcripts relative to wild-type were identified. The reported variants include point substitutions (K163A, K164A, K172G, K172L, R173C), deletions (del(K163,K164), del(K172,R173)), insertion mutants (insG180, insG181), and larger substitution blocks (subs(163--169)$\rightarrow$A; subs(178--193)$\rightarrow$WIHM). Given their localization within a region that shapes RNA exit and RNA–protein contacts, these mutations likely decrease the probability of forming unstable or termination-prone EC, thereby favoring readthrough over premature release \cite{lyakhov1997}.

A major advance in understanding the relationship between transcription termination and dsRNA formation was recently reported by \cite{he2026}, who demonstrated that the RNA exit channel of T7~RNAP acts as a central structural determinant of both processes. 
Through systematic mutagenesis of residues lining the RNA exit tunnel, including substitutions and deletions in subdomain~H (N171G, K172G, R173G, V174G, M183E, and multiple deletions spanning residues 167–174), as well as mutations in the thumb subdomain and specificity loop (K389G and N754G), the authors showed that increasing the effective diameter of the exit channel markedly enhances 3'-terminus homogeneity in run-off RNA products and reduces dsRNA formation.
Importantly, these effects were independent of the specific residue type and instead correlated with the minimum distance between the exit tunnel and the nascent RNA, highlighting a predominantly steric mechanism. 

Functionally, mutations that enlarge the RNA exit tunnel simultaneously decreased the efficiency of class~I T7 terminators, which are dependent on the formation of specific RNA structures, and reduced the formation of 3'-extended RNA species. Class~I termination relies on steric clashes between RNA hairpin structures and the RNA exit channel, which promote transcript shearing and termination. Mutations that widen the tunnel may alleviate these steric clashes, thereby reducing termination efficiency. Moreover, tunnel enlargement facilitates proper RNA exit, reducing 3'-extension of run-off transcripts and, consequently,  dsRNA formation \cite{he2026}.

Notably, combining the identified exit-tunnel-enlarging mutation M183E with I210V, a substitution located within the promoter-binding domain that plays a crucial role in promoter recognition and binding, resulted in a remarkable reduction in dsRNA generation, to approximately 0.75\% of the wild-type level \cite{he2026}.

\paragraph{Termination attenuation via a more processive elongation complex.}
Using phage-assisted continuous evolution,  residue S43 has been identified as a key determinant of transcription termination efficiency \cite{wu2021}. S43 is located in the C-helix subdomain, within a loop that refolds into a helix during the IC--EC transition, and lies along the DNA template in the EC.
Substitutions S43L, S43K, S43D, S43Y, and S43F consistently attenuated termination across multiple intrinsic terminators, with S43Y showing the strongest effect \cite{wu2021}. The authors propose that bulkier residues at position~43 presumably make the EC more processive, rendering the polymerase less prone to pausing triggered by destabilized RNA:DNA hybrids, so lowering the efficiency of intrinsic termination \cite{wu2021}.
Notably, S43Y weakens RNA binding and thereby prevents efficient extension of RNA primer/template structures. This results in a loss of apparent RdRP activity of T7~RNAP, while leaving canonical DNA-dependent transcription largely unaffected \cite{wu2021}.

As mentioned in the previous Sections \ref{subsec:mutants_affinity} and \ref{subsec:mutants_ictoec}, S43 occupies a strategic position at the intersection of multiple engineering strategies targeting template selectivity, IC-to-EC transition dynamics, and termination attenuation \cite{wu2021, qin2026, dousis2023}. Located in the C-helix, this residue undergoes a conformational transition from loop to helix during the IC--EC shift, ultimately positioning itself along the DNA template within a solvent-inaccessible cavity of the NTD. Substitutions favoring bulkier residues at position 43 orchestrate a coordinated set of functional improvements: they (i) promote earlier adoption of the EC-like helical conformation, enhancing 3'-end homogeneity; (ii) increase EC processivity by raising the energetic barrier to premature termination; and (iii) reduce the enzyme's affinity for RNA substrates, disfavoring aberrant self-priming and reannealing. These mechanistically distinct effects converge on a common outcome: collectively suppressing the multiple pathways that generate by-products during transcription.

\paragraph{Reducing fragmentation by tuning RNA:DNA separation in the RNA exit channel.}
Whereas the studies above primarily frame the phenotype as altered intrinsic termination due to T7 terminators, fragmented RNA can also arise from inappropriate RNA:DNA separation during elongation. Residues in the RNA exit channel positioned within hydrogen-bonding distance of phosphates near the 5'-end of the nascent single-stranded RNA were targeted in \cite{he2024b}. These residues were proposed to influence how efficiently RNA is separated from the DNA template during elongation \cite{yin2002}. While efficient separation is necessary for processivity, overly rapid or poorly coordinated separation may promote pausing and destabilization, increasing the probability of releasing truncated RNA products.
Among tested variants, K389A produced the strongest improvement in RNA integrity by substantially reducing fragmented RNA by-products while preserving IVT yield relative to wild-type. Additional substitutions improved integrity more modestly: K389L/R/M in the thumb subdomain, N171A and K172A in the H~subdomain, and N754A in the specificity loop. In general, substitutions at these positions toward amino acids with higher $\alpha$-helix propensity tended to yield the largest integrity gains \cite{he2024b}. Interestingly, other mutations at the same positions discussed above were shown to widen the RNA exit channel, thereby reducing the efficiency of T7 terminators and the generation of 3'-extended by-products \cite{he2026}, highlighting the importance of these residues in interactions with  nascent RNA in the RNA exit channel.

\subsection{Immobilized or chimeric T7~RNAP} \label{subsec:mutants_immobilization}

A final class of approaches to reduce by-product formation does not involve tuning the intrinsic catalytic properties of T7~RNAP, but rather re-engineering the physical context in which transcription occurs. The shared objective is to suppress non-productive RNAP–RNA interactions (notably re-binding of run-off RNA than enables 3'-extension), while maintaining efficient promoter-driven initiation. Two conceptually related solutions have been explored to bias the enzyme toward DNA templates and away from RNA templates: (i) immobilization and co-tethering of T7~RNAP with its promoter DNA, and (ii) fusion of T7~RNAP with an additional DNA-binding domain.

\paragraph{Co-tethering T7~RNAP and promoter DNA to operate under high-salt conditions.}

High-salt IVT conditions (e.g., elevated NaCl) broadly destabilize electrostatic interactions between T7~RNAP and nucleic acids. This can be beneficial because it weakens re-binding of run-off RNA to the polymerase, thereby strongly suppressing the 3'-extension pathway. However, the same effect also compromises promoter binding and thus inhibits promoter-driven initiation, making high salt impractical for conventional solution-phase IVT \cite{cavac2021, malagodapathiranage2024}.

Co-tethering strategies resolve this apparent contradiction by selectively restoring promoter engagement through proximity: T7~RNAP and promoter DNA are physically tethered so that the local effective concentration of promoter DNA near the polymerase remains high even when bulk electrostatic interactions are weakened by salt. In this configuration, transcription can be performed at salt concentrations that suppress RNA re-binding and 3'-extension, while still enabling efficient synthesis of run-off transcripts, since elongation is comparatively resistant to high-salt conditions \cite{cavac2021, malagodapathiranage2024}. 

A practical way to implement co-tethering is through immobilization of T7~RNAP and the DNA template on a common solid support. Beyond maintaining promoter proximity, immobilization offers additional process advantages, including simplified separation of the RNA product from RNAP and template DNA at the end of the reaction and the potential reuse of the reaction components in subsequent transcription cycles \cite{cavac2021}.

In \cite{cavac2021}, the authors implemented a non-covalent co-localization strategy on magnetic beads. T7~RNAP was fused at its N-terminus to a Strep-tag II peptide (WSHPQFEK) with a short flexible GGS linker, and the 5'-end of the non-template DNA strand was biotinylated. Both Strep-tag II and biotin bind with high affinity to tetravalent Strep-TactinXT-coated magnetic beads, which immobilizes the polymerase and the promoter DNA on adjacent sites \cite{cavac2021}. Importantly, immobilization did not measurably impair catalytic activity. When co-tethered with promoter DNA, promoter-driven transcription was maintained at salt concentrations that strongly inhibit untethered transcription, while the 3'-extension activity of T7~RNAP was almost completely eliminated, improving both yield and product purity \cite{cavac2021}. A limitation noted in this system is that, based on preliminary observations, DNA longer than 120~bp may not be immobilized efficiently \cite{cavac2021}.

A recent study addressed this limitation by using a covalent tethering architecture \cite{malagodapathiranage2024}. In this design, T7~RNAP is fused at its N-terminus to a HaloTag domain, which can be covalently crosslinked to a chloroalkyl-modified promoter DNA duplex positioned near the promoter-binding site of the polymerase. The downstream 5'-end of the DNA carries a biotin, enabling immobilization of the entire polymerase–promoter complex on streptavidin-coated magnetic beads. This co-tethered system again displayed higher salt tolerance than uncoupled species and, under high-salt IVT conditions, enabled production of high-quality mRNA with minimal to negligible dsRNA impurities. Notably, it also required substantially lower amounts of DNA and RNAP to reach a given yield output \cite{malagodapathiranage2024}.

\paragraph{Biasing template choice through fusion with DNA-binding domains.}

The fusion of T7~RNAP with DNA-binding domains has been explored in recent years to generate programmable T7-based synthetic transcription factors \cite{hussey2018programmable,lee2026t7}. This strategy is also a conceptually related alternative to the co-tethering of T7~RNAP and promoter DNA for improving IVT purity, presented in the previous subsection. Indeed, an additional DNA-binding domain increases the likelihood of productive DNA-templated initiation and elongation while reducing interactions with RNA templates that promote RNA-dependent synthesis and dsRNA formation.

A chimeric T7~RNAP was engineered by fusing an archaeal DNA-binding domain to the N-terminus of the polymerase using a flexible 30-amino acid linker in \cite{sanjeev2025}. The resulting chimera produced substantially less dsRNA (reported as a three- to fourfold reduction relative to wild-type) and exhibited improved salt tolerance \cite{sanjeev2025}. In practice, combining such a template-biasing chimera with high-salt IVT conditions creates a favorable regime for suppressing dsRNA generation and can reduce or eliminate the need for additional downstream purification \cite{sanjeev2025}. 

A similar strategy was also explored in \cite{ong2021}. T7~RNAP was fused to several heterologous DNA-binding domains, including the Sso7d DNA-binding protein, a helix–turn–helix DNA-binding domain from \emph{Pyrococcus furiosus}, and a LacI-like DNA-binding domain from \emph{Thermotoga}. Among the variants tested, an 18-mutation T7~RNAP fused to the \emph{Thermotoga} LacI-like DNA-binding domain exhibited markedly improved activity at elevated temperatures together with a reduction in transcriptional by-products.

\begin{table}[p]
\centering
\small
\setlength{\tabcolsep}{5pt}
\renewcommand{\arraystretch}{1.25}
\caption{Summary of T7~RNAP engineering strategies to reduce by-product formation during \textit{in vitro} transcription.}
\label{tab:T7RNAP_strategies}
\begin{tabular}{p{2.5cm} p{3.25cm} p{3.25cm} p{2.6cm} p{3.6cm}}
\toprule

\textbf{Strategy} & \textbf{Target Mechanisms} & \textbf{Representative Substitutions} & \textbf{Reduced   \hspace{1cm}by-product} & \textbf{Key trade-offs and\hspace{1cm} remarks} \\
\toprule

\textbf{Increasing RNAP thermostability} 
& Reduced RNA-template affinity and RNA secondary structure stability at elevated temperatures 
& S430P, N433T, C510R, S633P, F849I, F880Y
& 3'-extended RNAs; RNA-templated antisense RNAs
& RNA degradation risk at high temperature; terminal transferase activity persists; most effective with poly(A)-encoded templates \\
\midrule

\textbf{Reducing\newline the affinity for\newline non-specific\newline templates} 
& Suppression of RNA-templated synthesis and promoter-independent initiation 
& S43E, G47W, A70Q, F162S, K182E, R386F/Y/W, R632N, S633P, Q649L, G753A, Q786M, F880A, 884G
& 3'-extended RNAs; both types of antisense RNAs
& Often reduces dsRNA strongly; may reduce yield or RNA integrity\\
\midrule

\textbf{Smoothing the \newline IC--EC transition} 
& Reduced abortive\newline cycling and SAT release during initiation 
& G47A, A70Q, M183E, I210V, P266L
& Abortive RNAs 
& Primarily reduces SATs production; context-dependent indirect reduction of dsRNA \\
\midrule

\textbf{Reducing\newline premature\newline termination} 
& Stabilization of\newline RNA exit channel or\newline more processive EC
& S43Y, N171A/G, K172A/G, M183E, K389A/G, N754A/G
& Truncated RNA
& Primarily improves RNA integrity; reduction of 3'-extended or RNA-templated antisense transcripts in some cases \\
\midrule

\textbf{Immobilized or\newline chimeric RNAP} 
& Preserving DNA-templated transcription while disfavoring RNAP–RNA template interactions under high-salt conditions
& RNAP\newline co-tethered with promoter DNA or fused with additional DNA-binding domains
& 3'-extended RNAs; RNA-templated antisense RNAs 
& Very strong dsRNA suppression; increased process complexity; scalability and materials considerations; enables RNAP/DNA reuse \\
\bottomrule
\end{tabular}
\end{table}


\section{Tackling trade-offs in T7~RNAP engineering}

The studies reviewed here collectively illustrate that by-product formation during \textit{in vitro} transcription is not driven by a single aberrant activity of T7~RNAP, but rather emerges from the intrinsic coupling between its conformational dynamics, substrate recognition, and catalytic efficiency. As summarized in Table~\ref{tab:T7RNAP_strategies}, engineering strategies that reduce dsRNA, short abortive transcripts, or truncated RNA products target distinct mechanistic layers of the transcription cycle, including template selection, conformational dynamics during initiation, elongation stability, termination behavior, and reaction geometry. Importantly, these layers are highly interconnected, such that perturbations intended to suppress non-canonical activities often propagate to core aspects of transcriptional performance.

A recurring theme across T7~RNAP engineering efforts is therefore the intrinsic trade-off between transcriptional specificity and enzymatic processivity. Mutations that suppress RNA-dependent RNA synthesis, aberrant extension, or premature termination frequently alter conformational flexibility, substrate binding, or initiation dynamics, which can in turn compromise overall transcription efficiency. This coupling reflects the unusually low intrinsic stability and high conformational plasticity of wild-type T7~RNAP, which are properties  essential for rapid promoter clearance and highly processive elongation, yet simultaneously permit non-canonical activities such as RNA-templated synthesis, loopback extension, and promoter-independent initiation.

Recent studies increasingly demonstrate that this trade-off is not absolute and can be mitigated through rational combinations of mutations with complementary effects. For example:
\begin{itemize}

    \item The G753A substitution reduces antisense transcription but also increases RNA fragmentation, which motivated its combination with K389A, a mutation reported to improve RNA integrity. This combination reduced dsRNA levels while restoring RNA integrity without compromising RNA yield or protein expression \cite{he2024, he2024b}.

    \item A multi-site variant was constructed by combining substitutions in the N- and C-terminal cavities known to disfavor non-specific template engagement (S43E, S633P, and Q786M), thermostabilizing substitutions (S397W and S430P), and Q744R reported to compensate for the loss of activity associated with the thermostabilizing mutations. This engineered variant achieved a tenfold reduction in dsRNA production at~37°C and improved thermostability, while promoting canonical DNA-templated transcription over RNA-templated side activity \cite{qin2026}.

    \item Combining G47A, which reduces abortive cycling, with the C-terminal 884G addition, which reduces RNA-template affinity, produced synergistic improvements with a marked reduction in dsRNA content without compromising RNA yield \cite{dousis2023}. This example highlights how EC stabilization can complement template-biasing strategies. Similarly, the combination of the F162S and K180E substitutions, which reduce ssRNA binding, with A70Q, which stabilizes the EC conformation, has also been shown to improve T7 RNAP performance \cite{tang2025}.

    \item The double mutant M183E/I210V acts on two distinct mechanisms of dsRNA formation: M183E reduces premature termination and 3'-extension of transcripts, while I210V reduces abortive cycling. dsRNA levels are reduced by more than 99\% relative to wild-type T7~RNAP, without detectable loss of transcriptional yield or translational efficiency \cite{he2026}. The performance of this double mutant illustrates how the combination of initiation- and elongation-stage modifications can overcome the trade-offs typically associated with single-mechanism mutations.
    
\end{itemize}
These few examples emphasize that robust suppression of transcription by-products generally requires a combination of complementary strategies. Conversely, strategies that act on orthogonal axes, such as immobilized or chimeric T7~RNAP, can suppress by-products without directly affecting the enzyme's activity, thereby not requiring mitigation by other mutations.

\bigskip
An important implication of these findings is that the performance of engineered T7~RNAP variants is highly context-dependent. The impact of a given mutation on by-product formation, transcriptional efficiency, and specific activity depends not only on the intrinsic properties of the enzyme but also on template architecture (e.g., poly(A)-encoded versus enzymatically added tails), reaction conditions (temperature, ionic strength, magnesium concentration), and the intended RNA modality.
Crucially, biochemical performance must be evaluated alongside practical and economic considerations: a variant that strongly suppresses dsRNA or improves fidelity may nonetheless require substantially higher enzyme loading or elevated substrate concentrations to achieve comparable yield and purity. In many manufacturing contexts, a mutant that reduces dsRNA but requires several-fold more polymerase would be impractical. However, certain specialized applications may prioritize reaction quality over cost or throughput, making such trade-offs acceptable. Overall, assessing engineered RNAPs requires integrating biochemical performance with resource consumption, purification burden, and the specific constraints of the intended applications.

These observations further highlight that enzyme engineering and process engineering represent complementary, rather than competing, routes to improving IVT product quality. Strategies such as high-temperature transcription, high-salt conditions combined with RNAP–promoter tethering, or spatial confinement of transcription components directly reshape the reaction environment in ways that suppress non-canonical RNAP activities while maintaining favorable specific activity. In several cases, these approaches achieve levels of dsRNA suppression comparable to or exceeding those obtained through active-site mutations, and they do so without imposing large increases in enzyme or substrate consumption. This suggests that optimal IVT platforms will likely integrate tailored polymerase variants with reaction conditions explicitly chosen to minimize by-product formation while remaining compatible with the economic and operational constraints of the target application.

Taken together, the emerging picture is that robust suppression of transcriptional by-products will require multi-parameter optimization across enzyme sequence, reaction conditions, and template design. Rather than converging toward a single “ideal” polymerase, future IVT systems may rely on modular RNAP variants and process configurations tuned for specific RNA formats, manufacturing constraints, and cost-quality trade-offs. Beyond classical mutagenesis, chimerization offers an additional route for functional diversification---enabling incorporation of domains that support IVT-relevant activities such as co-transcriptional capping, pyrophosphatase activity, or other accessory functions that decouple productive transcription from the side activities inherent to T7~RNAP. Early demonstrations of such hybrid architectures suggest that expanding the functional repertoire of single-subunit polymerases may become an important direction for next-generation IVT systems.


\printbibliography

\end{document}